\documentclass[twocolumn,prl]{revtex4}
\usepackage{amsmath}
\usepackage{amssymb}
\usepackage{latexsym}
\usepackage{graphicx}% Include figure files
\usepackage{gensymb}
\usepackage{bm}% bold math

\def\cP{{\cal P}}

\def\eqnn#1{Eq.~(\ref{eq:#1})}

\def\figno#1{Fig.~\ref{fig:#1}}

\def\Im{{\rm \,Im}}

\def\vev#1{\left\langle #1\right\rangle}
\def\Im{{\rm \,Im}}

\def\kb{k_{\scriptscriptstyle\rm B}}

\def\figW{80mm}
\begin{document}
\title{
 Observing individual thermal motions of ions, and  molecules in water with light}
\author{Kenichiro Aoki\footnote{E--mail:~{\tt
      ken@phys-h.keio.ac.jp}.}, and Takahisa
  Mitsui\footnote{E--mail:~{\tt mitsui@phys-h.keio.ac.jp}.}  }
\affiliation{Research and Education Center for Natural Sciences and
  Dept. of Physics, Hiyoshi, Keio University, Yokohama 223--8521,
  Japan}
\begin{abstract}
  We observe thermal motions of ions, and molecules in water through
  light extinction, at the individual particle level.  The motions
  appear as time dependent intensity variations, characterized through
  their averaged spectra.  Theoretical spectrum derived from random motions of
  one particle describes these measured spectra. This theory is used to
  extract diffusion constants of liquid mixtures and solutions, that
  correspond to binary diffusion, and thermal diffusion, which are
  consistent with previous macroscopic measurements. We also estimate
  the sizes of the particles.
\end{abstract}
    \maketitle 
\par
All matter gives rise to thermal motions at the molecular level, which
can be observed, for instance, indirectly through Brownian motion.
Diffusion manifests thermal motions macroscopically, and is important
not only to physics, but also to chemistry, biology, technology, as
well as culinary science\cite{Cussler}.  Being able to ``see'' the
motion of individual particles directly, is not only fascinating, 
but should contribute to a better understanding of the microscopic
structure that underlies macroscopic diffusion.
Previously, nucleation processes\cite{nucleation1,nucleation2}, and
motions of gold atoms in ion liquid\cite{Miyata2017} were observed at
the atomic level, using electron tomography.

In the present work, we shine light through liquids, and observe the
thermal motions of individual ethanol molecules, and NaCl
ions in water, through the intensity variations in the transmitted
light they cause, for the first time. This method is
referred to as transmission fluctuation spectroscopy (TFS), below. The
thermal motions are characterized through the averaged spectra of light
intensity variations, and using the theory of random motions of
particles, we relate the properties of these spectra to the
macroscopic diffusive behavior of NaCl, ethanol, and water.
TFS was previously applied to rubidium gas, where averaged random
motions of individual atoms were observed through their
shadows, using resonant extinction\cite{rbWalk}.
Our intent here is to study ``everyday'' substances, at room
temperature, and pressure.  Such properties of of NaCl solutions, and
ethanol-water mixtures are of interest to a broad range of fields, and 
belong to  a classic, yet active  area of
research.\cite{water1,saltWater1,saltWater2,alcWater1,alcWater3}.

TFS is complementary to dynamic light scattering (DLS),
which has been, and is a powerful method to study properties of
matter\cite{DLS}. In DLS, one usually studies light scattered by matter, away
from the forward direction. In TFS, one studies light intensity
variations in the forward direction, caused by light extinction in its
path.
In TFS, since the vast majority of the light is neither scattered nor absorbed,
shot-noise, often referred to as the standard quantum limit (SQL), can
dominate the intensity variations, if left as is. This noise needs to be
substantially reduced to measure minuscule variations caused by light
extinction due to ions, and molecules in water, especially since the
extinction is {\it non-resonant} for the materials we study.

\begin{figure}[htbp]
  \centering \includegraphics[width=\figW,clip=true]{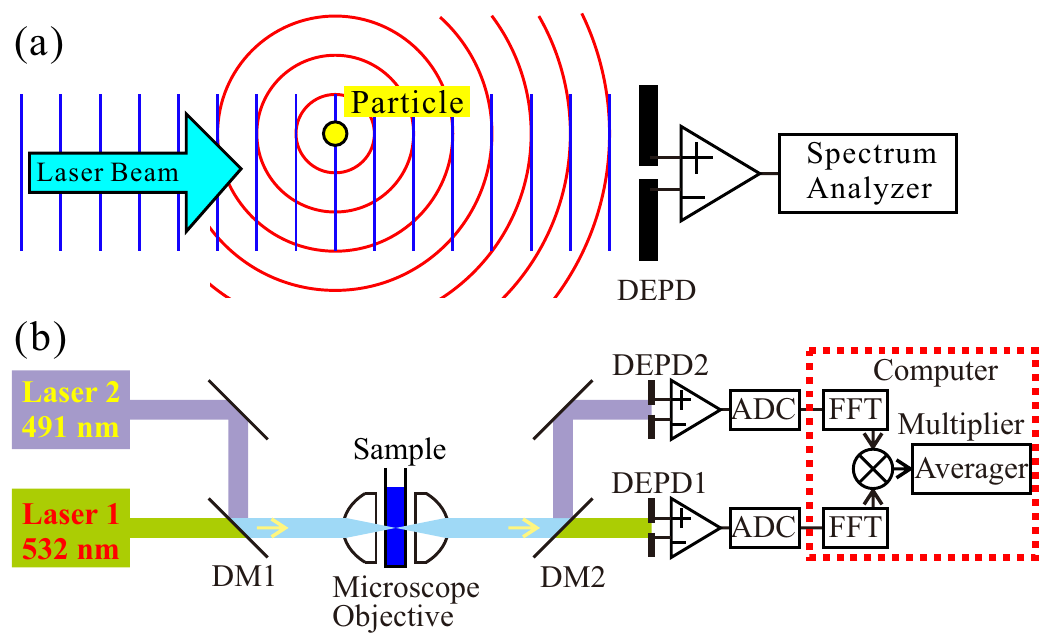}
  \caption{(a) The principle underlying the experiment. (b) The
    schematic of the experimental setup: Light beams from two light
    sources (wavelengths 515\,nm, 532nm) are merged using a dichroic
    mirror (DM1), focused with a microscope objective, and shone on
    the sample. The power of the beams at the sample are 550$\,\mu $W,
    870$\,\mu $W, respectively. The light in the forward direction is
    collimated using another microscope objective, split into the two
    wavelength components via DM2, and fed into DEPD1, and DEPD2.  The
    differences in the photocurrents in DEPD1,2 are digitized through
    analog-to-digital converters (ADC), Fourier transformed (FFT), and
    their averaged correlation is computed on a computer. }
  \label{fig:setup}
\end{figure}
The principle underlying the experiment is illustrated in
\figno{setup}(a): Light extinction due to a particle in the light beam
changes the beam intensity that depends on the location of the
particle. In the experimental setup (\figno{setup}(b)),  laser
beam is shone on the fluid in a cell (depth 1\,mm), and the light in
the forward direction is detected by a dual element photo diode (DEPD,
S4204, Hamamatsu Photonics). The motion of a particle 
causes intensity variations which show up as time dependent
photocurrent differences in the two elements of the photodiode. The
spectrum is obtained by averaging, and Fourier transforming these
variations in the photocurrent.
We use two light sources with different wavelengths, and then take the
correlation of the intensity variation measurements from the two
DEPDs, corresponding to the two wavelengths, to statistically reduce
all uncorrelated noise in the two DEPD measurements\cite{MA1}.  This
principle is effective in reducing any uncorrelated noise in the two
DEPD measurements, including the shot-noise\cite{MA1,rbWalk,Pottier17}. The
magnitude of the spectrum is normalized by the shot-noise level
determined from the measured photocurrents.

The thermal motions of one particle in a fluid can be modeled by a
random walk process.
Using a theory similar to that for 
the transmission intensity variations caused by one particle derived in
\cite{rbWalk}, the spectrum corresponding to our current experiment is
\begin{equation}
  \label{eq:spec}
  S(f) = {\vev{|\Delta \tilde \cP|^2}\over \cP^2} = AF(\pi w^2f/D)\ .
\end{equation}
Here, $\cP$, $\Delta \cP$ are the (average) power of the light beam at
the two elements of DEPD, their difference, and tilde denotes the
Fourier transform. $D$ is the diffusion constant, $w$ the beam radius,
and $f$ the frequency.
$ F(x)=2-2\Im\left[xe^{ix}{\rm Ei}(-ix)\right]$, where ${\rm Ei}(x)$
is the exponential integral function\cite{Gradshteyn}. It is important
to note that the shape of the spectrum is uniquely determined by the two
parameters, $D$, and $w$.

\begin{figure}[htbp]
  \centering \includegraphics[width=\figW,clip=true]{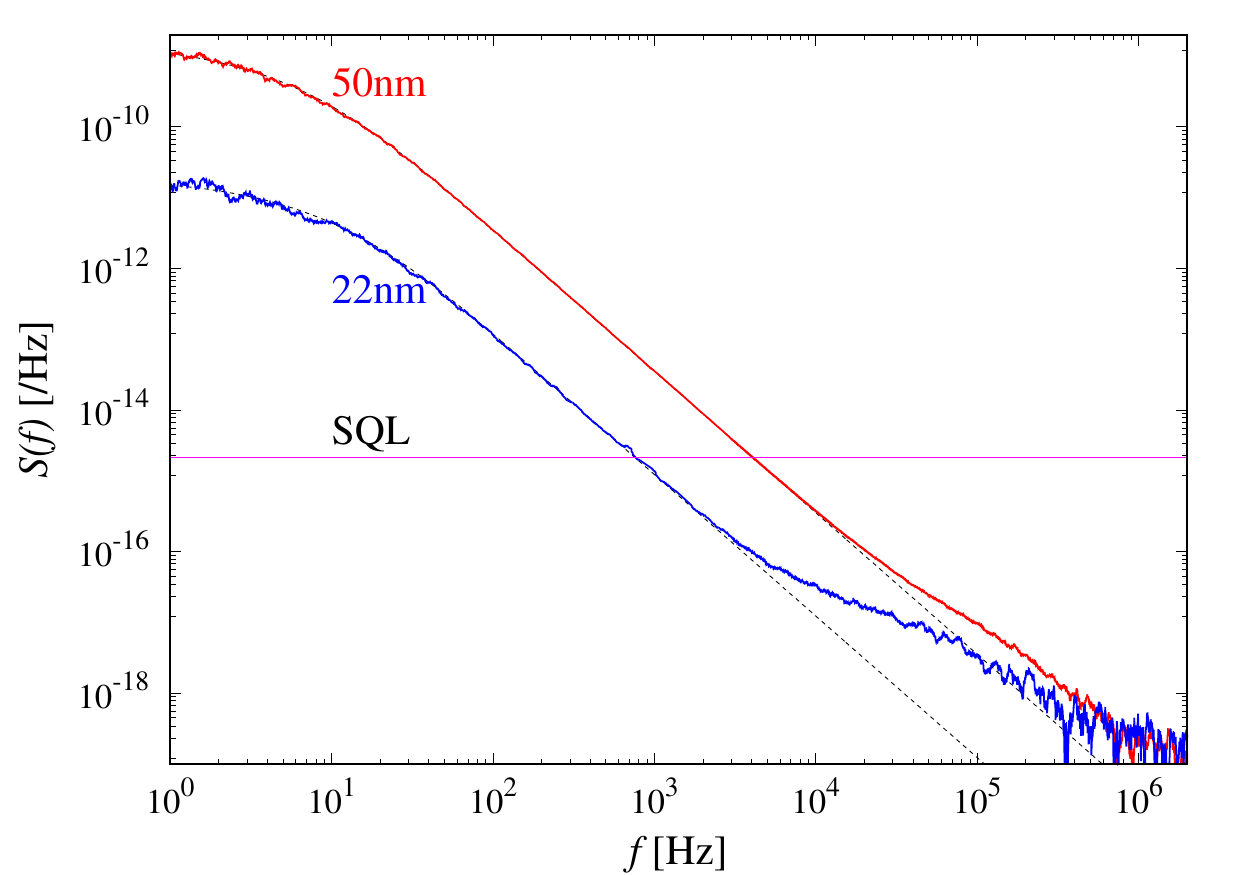}
  \caption{TFS spectra for polystyrene beads in water ($a=25\,$nm,
    $n=7.3\times10^{16}\,\rm m^{-3}$, $t=24.5$\degree C) (red), 
    ($a=11\,$nm, $n=3.6\times10^{17}\,\rm m^{-3}$, $t=24.5$\degree C) 
    (blue).
    Corresponding theoretical spectra \eqnn{spec} (black, dashed) are seen to
    agree well with the experimental results. SQL is also indicated. }
  \label{fig:poly}
\end{figure}
In \figno{poly}, the measured TFS spectra for polystyrene beads in
water, with radii 25\,nm (Polysciences, Inc.)
, and 11\,nm (Thermo Fisher Scientific) 
are shown. The difference in the index of refraction for polystyrene,
and water causes light extinction, leading to the observed intensity
fluctuations due to Brownian motion. The Einstein relation
$D=\kb T/(6\pi\eta a)$, ($\eta$: water viscosity\cite{CRC}, $a$:
particle radius) lets us compute the diffusion constant, which was
used in the theoretical spectra, \eqnn{spec} in \figno{poly}. Here,
effective $w$ was deduced from the experimental spectra. The spectral
shapes match well between the theory, and the experiment, over six
orders of magnitude.  From SQL indicated in \figno{poly}, it can be
seen that such dynamic range is achieved only by using the reduction
of shot-noise.  $w$ estimated from the two spectra are consistent with
each other, and also with the design of the experiment.  The
magnitudes of the spectra were obtained from the experimental spectra.
These spectral measurements of particles of known size provide us with
a method to measure $w$, whose precise value can vary slightly under
the experimental conditions, and is also difficult to measure
directly.  In the subsequent measurements of other samples, $w$ is
determined from TFS spectra of $a=25\,$nm polystyrene beads, prior to,
and after the sample measurement, where $w$ varied from $0.99$ to
$1.07\, \mu$m.  There is a mismatch between the theoretical, and the
experimental spectra below the magnitude $10^{-16}$\,/Hz, due to
temperature fluctuations of water, as explained below.

\begin{figure}[htbp]
  \centering \includegraphics[width=\figW,clip=true]{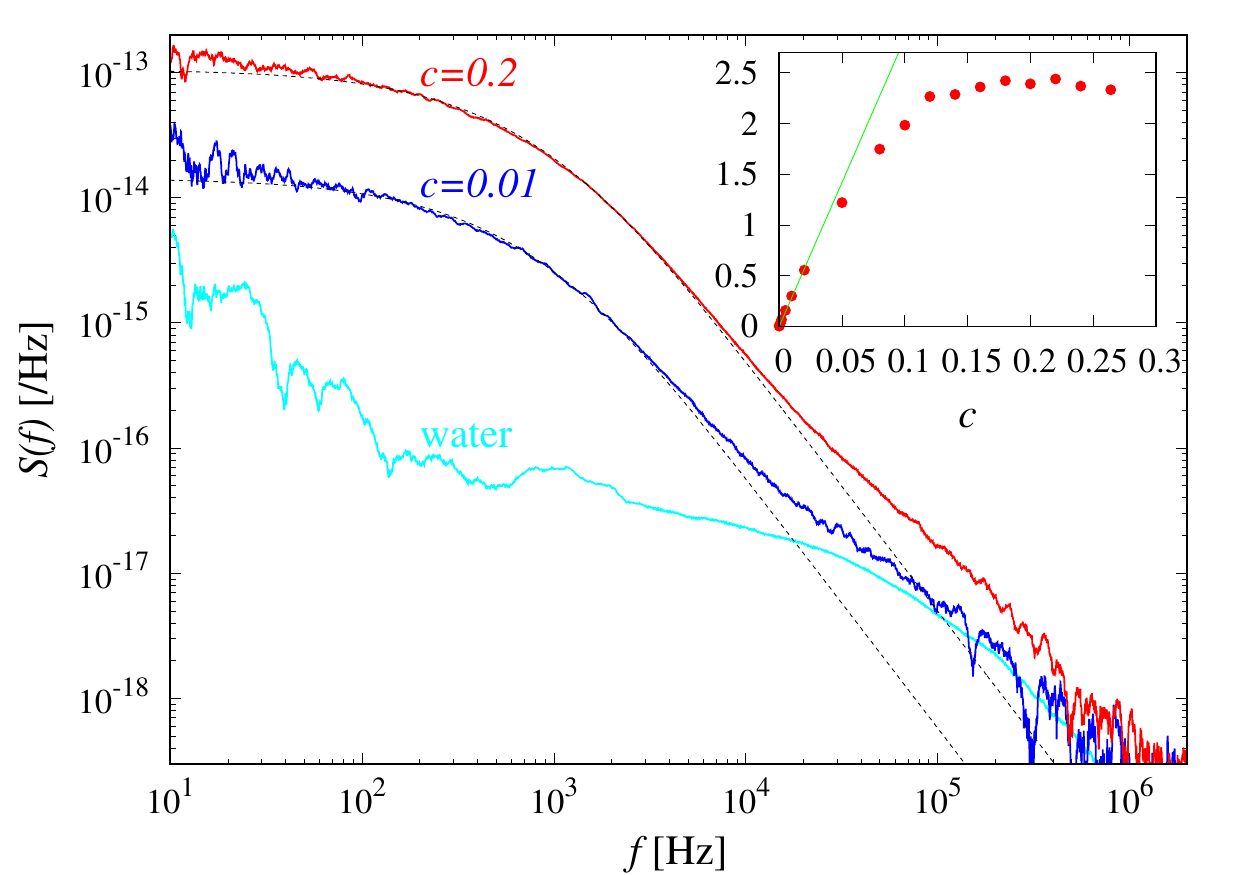}
  \caption{TFS spectra for the aqueous solution of NaCl at
    mass fractions, $c=0.2$ (red), $c=0.01$ (blue), and water (cyan),
    at 24.5\degree C. TFS spectrum is seen to clearly differ from that
    of water, even at $c=0.01$.  Corresponding theoretical spectra for
    diffusion of NaCl diffusion in water, and (black,
    dashed), and temperature fluctuations (black), are also shown.  (Inset)
    The dependence of the spectral magnitude at 300\,Hz
    ([$10^{-14}$/Hz]) on the mass fraction of NaCl, $c$.
    Linear fit for lower concentrations is also shown.}
  \label{fig:nacl}
\end{figure}
In \figno{nacl}, the experimentally measured TFS spectra are shown for
aqueous solutions of NaCl, with mass fractions, $c=0.2, 0.01$, and for
water.  The theoretical spectrum, \eqnn{spec}, is seen to describe the
measured spectra well for $f\lesssim 10^4$\,Hz. For
$f\gtrsim10^4$\,Hz, the temperature fluctuations of water also
contributes. Since $w$ is known from the polystyrene bead spectra, we
can deduce the only remaining parameter determining the spectral
shape, the diffusion constant, $D$, corresponding to each spectrum.
Binary diffusion spectra are pronounced at around $f=300$\,Hz, and the
spectral magnitudes there increases linearly with the concentration,
reflecting linearity with respect to the number of particles, and
saturates at higher concentrations (\figno{nacl} Inset).

\begin{figure}[htbp]
  \centering \includegraphics[width=\figW,clip=true]{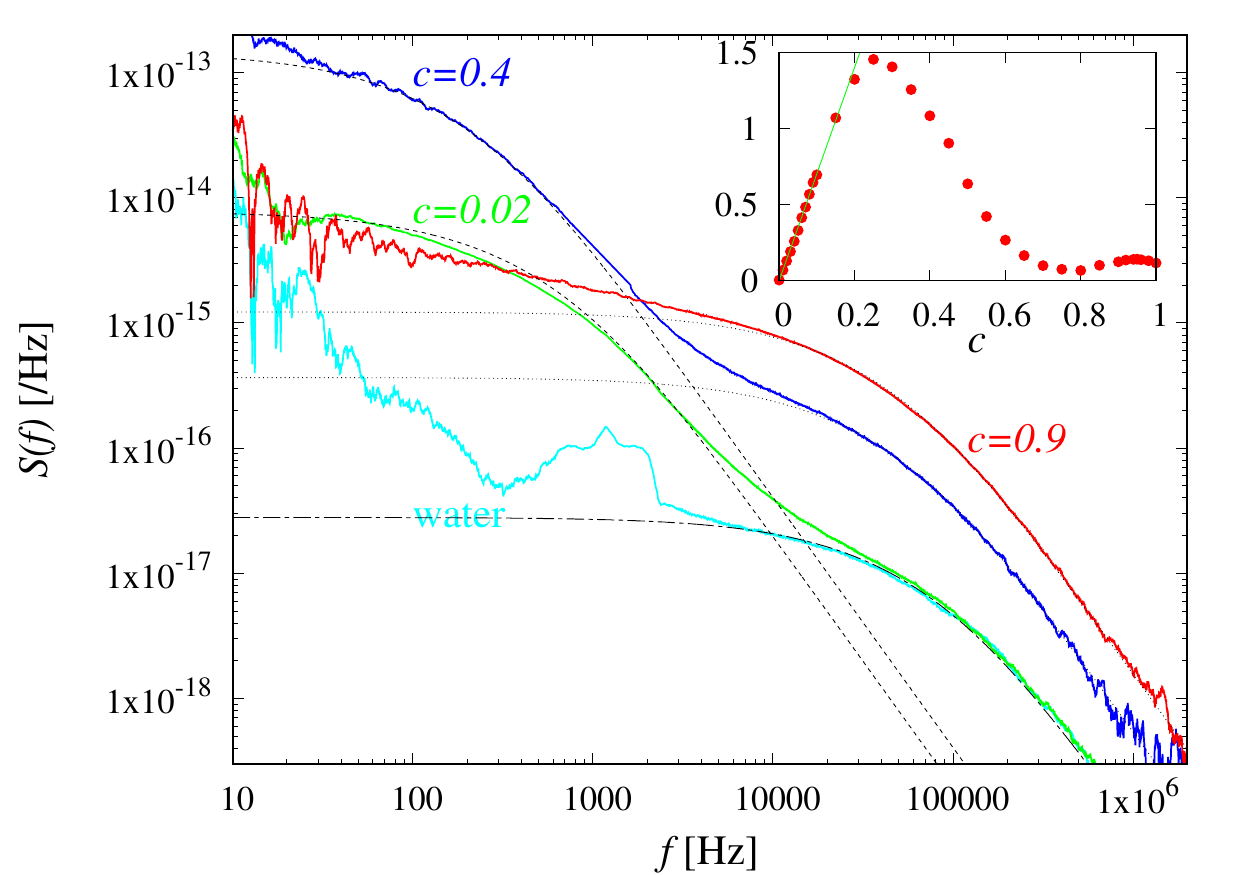}
  \caption{TFS spectra for ethanol--water mixture, with mass fractions
    $c=0.9$ (red), $c=0.4$ (blue), $c=0.02$ (green), and water (cyan),
    at 24.5\degree C.  Presence of ethanol is clearly visible even at
    $c=0.02$.  Corresponding theoretical spectra for diffusion of
    ethanol in water (black, dashed), and temperature fluctuations of ethanol
    (black, dotted), and water (black, dot-dashed) are also shown.  (Inset)
    The dependence of the spectral magnitude  on $c$ at 300\,Hz
    ([$10^{-14}$/Hz]).  Linear fit for lower concentrations is
    also shown.}
  \label{fig:ethanol}
\end{figure}
TFS spectra for ethanol are shown in \figno{ethanol} at $c=0.9, 0.4,0.02$,
along with that for water.  The theoretical spectrum agree
well with the measured spectra, for $f\lesssim 10^3$\,Hz, from which
$D$ can be computed. For $f\gtrsim10^4\,$Hz, the spectrum due to
temperature fluctuations in water becomes dominant, and its
theoretical spectrum, discussed below, is seen to describe the
experimental spectrum well. The spectral magnitude at $f=300$\,Hz
(\figno{nacl} Inset) exhibits intriguing behavior; for $c\lesssim0.2$,
the behavior is linear with respect to $c$, similarly to NaCl solution
above.
In the high concentration limit, there is no binary diffusion, and
the corresponding spectrum should decrease, which explains this
overall behavior. There is interesting non-monotonous behavior at
$c\simeq0.95$, which warrants further investigation.

\begin{figure}[htbp]
  \centering \includegraphics[width=\figW,clip=true]{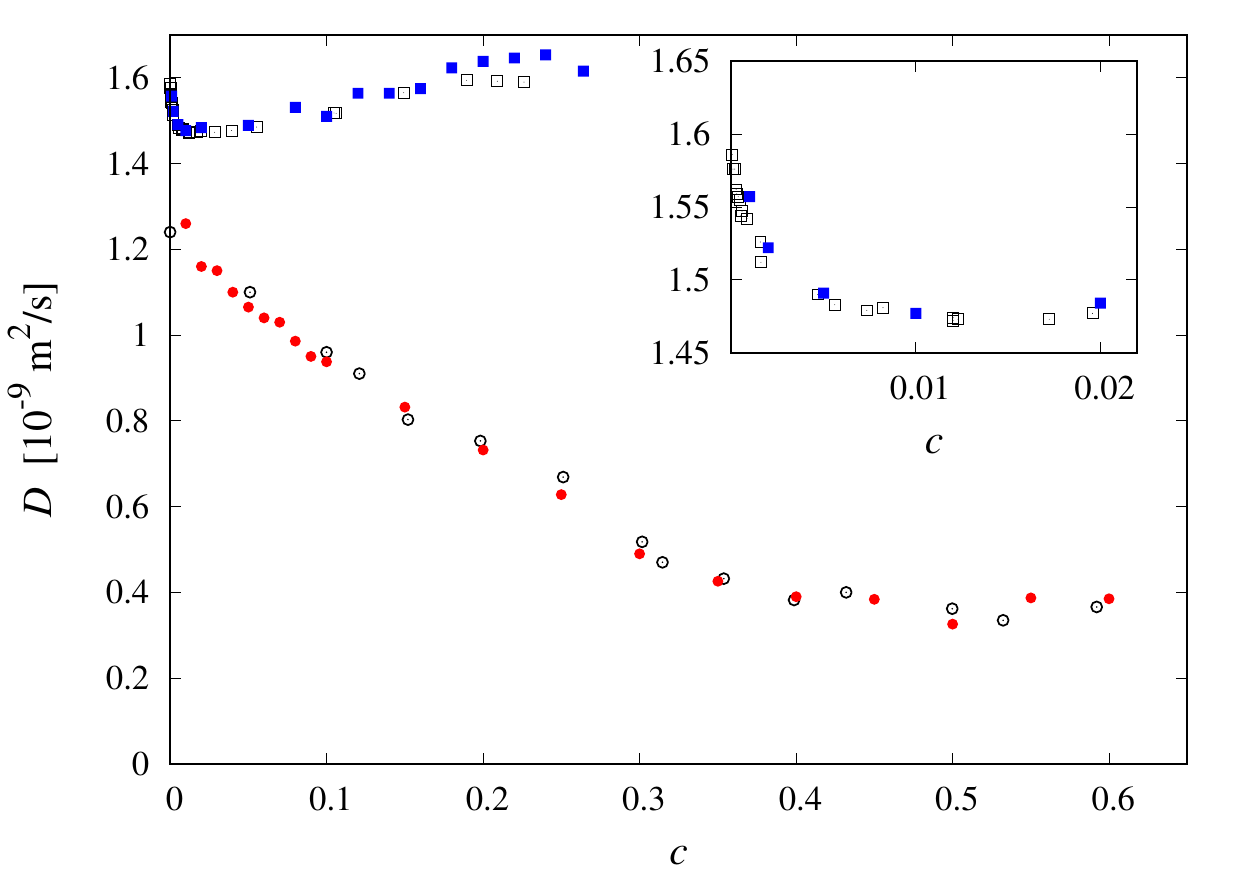}
  \caption{Diffusion constants obtained from TFS spectra, for NaCl
    solutions (\rule{0.5em}{0.5em}), and ethanol--water mixtures
    ($\bullet$) at various concentrations at 24.5\degree C.  The
    corresponding values in the previous literature for 25\degree C
    are also shown ($\Box, \ \circ$, respectively). Inset: Small $c$
    region for NaCl solutions.}
  \label{fig:concDep}
\end{figure}
The measured diffusion constants from TFS spectra for NaCl solutions,
and ethanol--water mixtures at various concentrations are shown in
\figno{concDep}, along with their corresponding values in the previous
literature\cite{naclD1,naclD2,ethanolD1,ethanolD3}, which
are consistent with each other.  While it seems reasonable to expect
that the random thermal motions of molecules are directly responsible
for the diffusive behavior in liquid mixtures, and solutions, we find
it beautiful to see that this physics picture works so well.

\begin{figure}[htbp]
  \centering \includegraphics[width=\figW,clip=true]{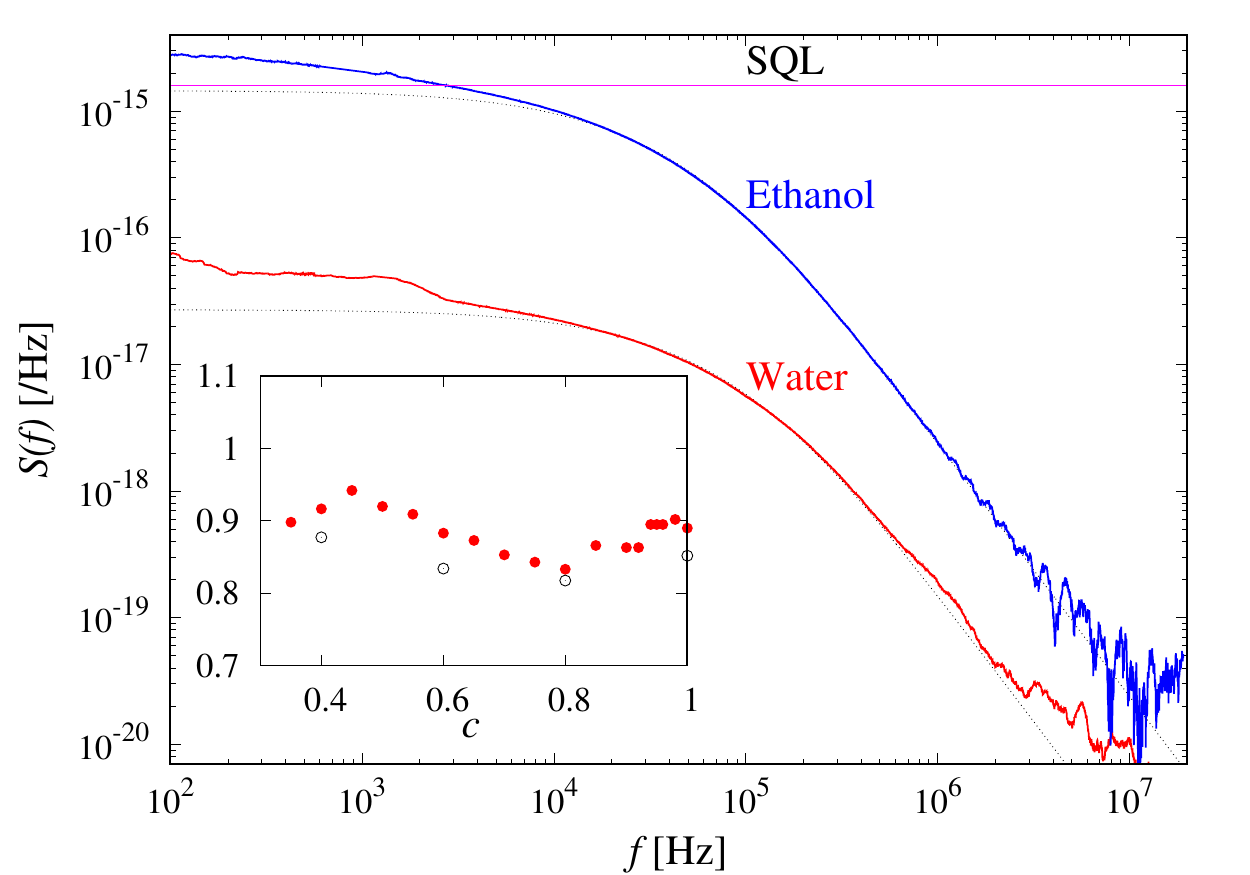}
  \caption{TFS spectra of ethanol (blue), and water (red), at
    24.0\degree C. Ethanol fluctuations are larger.  Corresponding
    theoretical spectra for temperature fluctuations (black, dotted)
    are seen to agree well with the experiment. SQL is also
    indicated. (Inset) Diffusion constants extracted from the spectra, $D$
    [$10^{-7}\rm \,m^2/s]$,  for ethanol-water mixtures at 24.5\degree C ($\bullet$), compared to
    the previously measured thermal diffusivity values, at 295\,K
    ($\circ$, line)\cite{ethanolTD1}.
  }
  \label{fig:thermal}
\end{figure}
In matter, there inevitably exist temperature and density
fluctuations, microscopically, that are responsible for Rayleigh
scattering, for instance\cite{LandauEM}.  These fluctuations should
give rise to local, and time varying thermal gradients, resulting in
{\it microscopic} thermal diffusive processes, in addition to binary
diffusion process discussed above.
Unlike binary diffusion, thermal diffusion does {\it not} necessarily
require mass diffusion, especially in liquids, at least at the
macroscopic level, since momentum transfer can occur without actual movement
at macroscopic scales.
The microscopic behavior underlying thermal diffusion in liquids is
not well understood, especially in mixtures\cite{Cussler,tDiffusion},
but the origin of these momentum transfer processes should lie in the
thermal motions of molecules.
Density fluctuations give rise to local, and time dependent
inhomogeneities in the susceptibility, causing light
extinction, even in pure liquids. A priori, it is
unclear whether these local thermal motions are directly observable,
and even if so, whether they reflect the properties of the thermal
diffusion process in a simple manner.
In studying TFS spectra of ethanol, and water, the spectra due to 
to the random thermal motion of molecules, \eqnn{spec}, are observed  
(\figno{thermal}). 
The diffusion constants extracted from the TFS spectra agree with
their corresponding values predicted from theory,
$\kappa/(\rho C_{\rm V})$ within few percent.  Here, $\kappa$, is the
thermal conductivity, $\rho$ the density, and $C_{\rm V}$ is the
specific heat\cite{CRC}. Furthermore, the diffusion constants were
measured at various ethanol concentrations in water, and compared to
previously measured values (albeit few degrees
different)\cite{ethanolTD1}, in \figno{thermal} (Inset), and 
agree within 10\%. 
Most of the measured spectra, or all for the case of
water, lie well below the shot-noise level, so that its reduction was
crucial for these measurements.

The magnitudes of TFS spectra reflect the cross sections of the
particles observed, and allow us to estimate the sizes of the
particles.  Here, we assume a simple power behavior with respect to
the particle radius, $a$, for the spectral magnitude per particle,
$A/n= Ca^p $ (\eqnn{spec}), where $n$ is the particle number density.
Using the size information of polystyrene beads, this leads to
$p=7.5$, and $a=0.30\,$nm, and $0.25$\,nm for NaCl (\figno{nacl}), and
ethanol (\figno{ethanol}), respectively.
While these estimates are crude, they have the merit of relying on a
simple relation, with no further assumptions.  The unit of motion in
diffusion, which has some hydration, is a topic of active research, and the
measured sizes are consistent with previous theories
\cite{saltWater2,alcWater3,molSize}.  
Studying this property in more detail through this new window is of
interest.
We note that the binary diffusion constant also leads to an estimate
of the size of the particle with the same order through the Einstein
relation, but the above estimate is optical, and is independent from
it, experimentally.
The power, $p$, is consistent with  extinction mainly due to
absorption, which should dominate for smaller particles\cite{Hulst}:
The cross sections due to absorption behave as $a^3$, and $A/n$ should
be proportional to its square. There is $1/D$ factor multiplying this
from the structure of the spectrum \eqnn{spec}\cite{rbWalk}, leading
$p=7$.  The value of $p$ found above is slightly larger, which is
consistent with some scattering contribution for larger beads, since
the relevant small particle expansion parameter is $ka=0.30$ for them
($k$: wave number), and cross sections behave as $a^6$ for Rayleigh
scattering\cite{Hulst}, leading to $p=13$.

In this work, we observed averaged thermal motions of ions, and
molecules that underlie binary diffusion, and thermal diffusion, as well as
thermal motions polystyrene particles, at the individual particle level.
This was made possible by TFS, which we propose as a method for the
detection of particle motions in fluids at the atomic scale.
We believe that being able to observe molecular motions at the atomic
scale complement previously available methods, and provides us with an
important window into the dynamical microscopic structure of liquids.
The binary diffusion constants, and thermal diffusivity deduced from
the TFS spectra agreed with those previously measured through
macroscopic methods. We obtained estimates of the sizes of ions, and
molecules in water, from the spectral magnitudes, that are consistent
with those obtained from other methods. We find the emergent picture
of the diffusive processes, from microscopic, to macroscopic quite
satisfying, and intriguing.

\end{document}